\documentstyle[twoside,fleqn,espcrc2,epsfig]{article}

\newcommand{\AmS}{{\protect\the\textfont2
  A\kern-.1667em\lower.5ex\hbox{M}\kern-.125emS}}
\def\frac#1#2{ {{#1} \over {#2} }}

\def\beq{\begin{equation}}
\def\eeq{\end{equation}}
\newcommand{\ba}         {\begin{eqnarray}}
\newcommand{\ea}         {\end  {eqnarray}}
\newcommand{\ban}        {\begin{eqnarray*}}
\newcommand{\ean}        {\end  {eqnarray*}}
\def\re#1{(\ref{#1})}

\def\np#1#2#3{Nucl.\ Phys.\ B#1 (19#3) #2}

\title{Developments and new applications of numerical stochastic 
perturbation theory}

\author{G. Burgio\address{Dipartimento di Fisica, Universit\`a di Parma 
	and INFN, Gruppo Collegato di Parma, Italy}, 
	F. Di Renzo\address{Department of Mathematical Sciences,
	University of Liverpool, United Kingdom}\thanks{Presenter of the 
		poster.}, 
	G. Marchesini\address{Dipartimento di Fisica, Universit\`a di Milano
	and INFN, Sezione di Milano, Italy}, 
        E. Onofri$^{\rm a}$, 
	M. Pepe$^{\rm c}$, 
	and L. Scorzato$^{\rm a}$} 

\begin{document}

\begin{abstract}
A review of new developments in numerical stochastic perturbation theory 
(NSPT) is presented. In particular, the status of the extension of the method
to gauge fixed lattice QCD is reviewed and a first application to compact 
(scalar) QED is presented. Lacking still a general proof of the convergence of 
the underlying stochastic processes, a self-consistent method for testing 
the results is discussed. 
\end{abstract}

\maketitle

\section{NSPT for gauge fixed lattice QCD}

Numerical stochastic perturbation theory (NSPT) was successful in performing 
high loops gauge invariant computations in lattice QCD \cite{8l} (for new, and 
potentially relevant, phenomenological implications see \cite{L2}). An 
extension to gauge fixed computations would of course be highly welcome. 
First results in this direction were reported in St. Louis, but the matter 
turned out to be not settled down. In the following, as a first step, the 
basics from \cite{mygf} are recalled. 

As it is well known, only gauge invariant quantities 
have a large time limit in standard stochastic quantization \cite{pw}, 
while gauge--non--invariant quantities are affected by divergences.  
As usual, the situation is slightly different on the lattice due to the 
fact that the gauge degrees of freedom have been compactified, so that 
gauge--non--invariant quantities average to zero. Gauge 
fixed Langevin simulations rely on the old observation \cite{gf0} that the 
standard evolution step 
\beq
U'_\mu(n) \, = \, e^{- F_\mu[\{U\},\eta](n)} \, U_\mu(n)
\eeq
($\eta$ is the gaussian noise)  can be interlaced with a gauge transformation 
\beq
\label{gt}
U^G_\mu(n) \, = \, G(n) \, U_\mu(n) \, G^\dag(n+\hat{\mu})
\eeq
Since 
\beq
F_\mu[\{U^G\},\eta] \, = \, G \; F_\mu[\{U\},G \eta G^\dag] 
\; G^\dag
\eeq
the convergence of the process for gauge invariant quantities is unaffected, 
while the gauge transformation $G$ can be chosen in order to enforce the 
gauge condition one is interested in: the stochastic process is then 
attracted towards the submanifold defined by the gauge condition and 
gauge--non--invariant quantities no longer average to zero. This recipe is in 
a sense the lattice implementation of Zwanziger's stochastic gauge fixing 
scheme \cite{zwa}.
 
NSPT rely on an expansion of the solution of Langevin equation in powers of 
the coupling, which actually decompactifies the fields (in a sense the 
fundamental fields of NSPT live in the Lie algebra), so that divergences in 
the gauge--non--invariant sector are back. Nevertheless one can adopt the 
stochastic gauge fixing scheme also in this context, with the caveat that 
also the gauge condition one wants to enforce has to be expanded as a series 
of conditions. To be definite, as the $A_\mu$ field is expanded as $A_\mu = 
A_\mu^{(0)} + g  A_\mu^{(1)} + g^2 A_\mu^{(2)} + \ldots$, the form of Landau 
condition one needs to impose is 
\beq
\label{land}
\partial_\mu A_\mu^{(k)} = 0
\eeq
for every order $k$ (partial derivatives, as usual, are to be understood as 
finite difference operators). \\
The implementation presented in \cite{mygf} goes as follows. One considers 
the functional (norm)
\beq
N[G] \, = \, \sum_{n,\mu} \mbox{Tr} ( A_\mu^G(n) {A_\mu^G}^\dag(n) ) \, = \, 
\sum_k g^k N^{(k)}
\eeq
where an expansion for $N$ is induced by the expansion of the field $A_\mu$. 
If the gauge transformation is chosen as
\beq
\label{oldgf}
G \, = \, e^{- \alpha w} \; \; \; \; \; \; 
w = \sum_k \, g^k \, \partial_\mu A_\mu^{(k)}
\eeq
extremum conditions for every $N^{(k)}$ are recovered that exactly enforce 
\re{land}. First simulations with this recipe were presented in \cite{mygf} 
and they seemed good up to order $g^4$. As a matter of fact, if one lets the 
system evolve for a while, the signal for gauge--non--invariant quantities 
is not at all stable, but one recovers a collection of plateaux. 
While gauge fixing transformation is in charge to kill 
divergences due to ``longitudinal'' degrees of freedom, \re{oldgf} lets 
some divergence still at work. This can be checked by inspecting the norms of 
the fields $A_\mu^{(k)}$. While the norm of $A_\mu^{(0)}$ is minimized and 
stable through the process (as a matter of fact, this norm is exactly 
$N^{(2)}$, which is the only $N^{(k)}$ one is ensured to minimize), higher 
orders suffer divergences in their norms. 

\begin{figure}[htb]
\vspace{9pt}
\begin{center}
\mbox{{\epsfig{figure=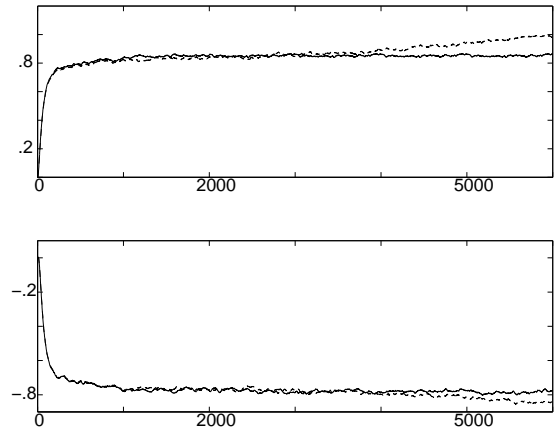,width=7.5 cm}}}
\end{center}
\caption{Above: the norm of the field $A_\mu^{(1)}$ with and without 
corrections to \re{oldgf}. Below: the signal for order $g^4$ of the trace of 
the link in the same simulations.}
\label{stab}
\end{figure}

A first attempt has been done in order to enforce \re{land} and keep 
every norm under control through the process. This can be obtained by 
adding new terms to $w^{(k)}$ in \re{oldgf} (the formulae are quite cumbersome 
and they will be written down elsewhere). The results are encouraging 
and can be summarized by inspecting fig.~\ref{stab}. In the upper half 
the norm of $A_\mu^{(1)}$ is monitored through the evolution with and 
without these new terms (\re{land} is enforced in either case). The lower 
half shows the signal for order $g^4$ of the trace of the link, again with and 
without these new terms: the new gauge fixing procedure keeps the plateau 
stable (while the old procedure lets the system depart from it).

\section{How can you trust NSPT?}

Of course a stable signal does not mean at all that a result is correct. 
Actually there is an embarrassing lack of a general proof of the convergence 
of the (quite cumbersome) stochastic processes underlying NSPT, so that in a 
sense one could worry about whether to trust a NSPT result or not. A general 
strategy to check a result can be outlined as follows: check up by explicit 
computation the trivial order and then set up relations among different 
perturbative orders. To be definite, one can think of Schwinger--Dyson 
equations as obtained by 
\beq
\label{sd}
\frac{1}{Z} \int D\phi \, \frac{\delta}{\delta\phi} \, ( O[\phi] 
\, e^{- S[\phi]} ) 
\, = \, 0
\eeq
By expanding this equation in the coupling one recovers relations among 
different orders. As it is well known, this is a standard approach to 
perturbation theory. In this context one wants to look at these relations 
to test whether they holds once every expectation value involved has been 
obtained by means of the Langevin technique. 

Unfortunately for (lattice) QCD \re{sd} reads 
\beq
\frac{1}{Z} \int DU \, \nabla ( O[U] e^{- S_W[U]} e^{- S_{GF}[U]} 
\Delta_{FP} ) \, = \, 0
\eeq
where $\nabla$ is the group derivative and $\Delta_{FP}$ the Fadeev--Popov 
determinant, which is tremendous to take into account without introducing 
ghosts. 

Nevertheless, from the standard formula for a QCD expectation value
\beq
\frac{1}{Z(g)} \int DA Dc D\overline{c} \; O[A] \; e^{- S[A]} \; 
e^{-S_c[A,c,\overline{c}]} 
\eeq
(where the action has been split in two terms $S$ and $S_c$, the latter 
being dependent on the ghosts), one can get a recursive relation for the 
coefficients of the expansion $\sum_k g^k O^{(k)}$. Namely, by deriving with 
respect to the coupling $g$, one gets ($O'$ means $d/dg \; O$)
\ba
(k+1) O^{(k+1)} &=& O'^{(k)} - (O S')^{(k)} \nonumber \\
&& + \sum_{l+m=k} S'^{(l)} O^{(m)} \\ 
&& + \sum_{l+m=k} {S_c}'^{(l)} O^{(m)} - (O {S_c}')^{(k)} \nonumber 
\ea
One can obtain by the Langevin technique everything except the last two 
contributions, which depend on the ghosts. For example, testing the order 
$g^4$ result for the trace of the link only requires to compute the 
ghost corrections to the gluon propagator.

\section{NSPT for compact QED}

A first implementation of the Schwinger--Dyson test program has been done 
within the context of a new application of NSPT, namely to compact QED. 
The final goal of this project is a collaboration with the Tor Vergata 
group, aiming at merging NSPT with the bermions approach \cite{berm} to push 
one loop further the computation of electron's $g - 2$. An important point 
to notice is that the extrapolation in the number of flavors is 
perturbatively well defined. Being QED not asymptotically free, a major 
problem to deal with is an explicit inclusion of counterterms. 

In order to gain experience, we started simulating scalar QED defined by 
the action 
$$ S = \frac{1}{e^2} \sum_P (1-\cos \phi_{\mu \nu}) - \sum_{n \mu} \phi^*(n) (U_\mu(n) \phi(n+\hat{\mu}) $$
$$+ U_\mu^\dag(n-\hat{\mu}) \phi(n-\hat{\mu}) ) 
+ \sum_n (M^2 + 8) \phi^*(n) \phi(n) $$
$\phi_{\mu \nu}$ being defined by the plaquette $U_{\mu \nu} = 
\exp{(i \phi_{\mu \nu})}$. 

We have till now performed computations up to order $\alpha^3$ in the gauge 
invariant sector, every result having been tested by Schwinger--Dyson 
equations. In many instances the overall picture is the same as for QCD; 
in particular, some form of stochastic gauge fixing is needed also in the 
gauge invariant sector in order to keep fluctuations under control.

\section{More to come \ldots}

Another application which is expected to come soon (some preliminary 
work has just started after the conference) is a high order perturbative 
expansion of the QCD running coupling $\alpha_{SF}$ as defined in the SF 
scheme \cite{lusc}. One is interested in pushing further the connection 
between different schemes and in evaluating the high order behavior itself 
of this definition of the coupling (in \cite{8l} the coupling $\alpha_{pl}$ 
defined via the plaquette has been found to be affected by a renormalon).


\end{document}